\documentstyle[preprint,prl,aps,epsf]{revtex} 
 
\begin{document}
 
\draft                    
 
 
\title{
         \vskip -0.5cm
         \hfill\hfil{\rm\normalsize Printed on \today}\\
         Fractional quantum conductance in carbon nanotubes }
 
\author{
         Stefano Sanvito,$^{1,2,3,}$\cite{{null},{Sanvitoemail}}
         Young-Kyun Kwon,$^{3}$
         David Tom\'anek,$^{3}$ and
         Colin J. Lambert$^{1}$
	 }
 
\address{$^{1}$ School of Physics and Chemistry, Lancaster University,
         Lancaster, LA1 4YB, UK}
 
\address{$^{2}$ DERA, Electronics Sector,
         Malvern, Worcs. WR14 3PS, UK}
 
\address{$^{3}$ Department of Physics and Astronomy, and
         Center for Fundamental Materials Research, \\
         Michigan State University,
         East Lansing, Michigan 48824-1116}
 
\date{Received \hspace{3.0cm} }
 
\maketitle
 
 
\begin{abstract}
Using a scattering technique based on a parametrized linear
combination of atomic orbitals Hamiltonian, we calculate the ballistic
quantum conductance of multi-wall carbon nanotubes. We find that
inter-wall interactions not only block some of the quantum conductance
channels, but also redistribute the current non-uniformly over the
individual tubes across the structure. Our results provide a natural
explanation for the unexpected integer and non-integer conductance
values reported for multi-wall nanotubes in Ref.\ \onlinecite{Heer98}.
\end{abstract}
 
 
 
Carbon nanotubes \cite{{Iijima91},{Dresselhaus96}} are narrow seamless
graphitic cylinders, which show an unusual combination of
nanometer-size diameter and millimeter-size length. This topology,
combined with the absence of defects on a macroscopic scale, gives
rise to uncommon electronic properties of individual single-wall
nanotubes \cite{{Iijima93},{Bethune93}}, which -- depending on their
diameter and chirality -- can be either conducting or insulating
\cite{{Mintmire92},{RSaito92},{Oshiyama92}}. Significant changes in
conductivity of these nano-wires may be induced by minute geometric
distortions \cite{Kane} or external fields \cite{Dekker-Nature98}.
More intriguing effects, ranging from the opening of a pseudo-gap at
$E_F$ \cite{{Louie98},{PRLxNTR}} to orientational melting
\cite{JMRxNTR}, have been predicted to occur when identical metallic
nanotubes are bundled together in the form of a ``rope''
\cite{Thess96}.
 
Electron transport in nanotubes is believed to be ballistic in nature,
implying the absence of inelastic scattering \cite{Heer98}. Recent
conductance measurements of multi-wall carbon nanotubes \cite{Heer98}
have raised a significant controversy due to the observation of
unexpected conductance values in apparent disagreement with
theoretical predictions. In these experiments, multi-wall carbon
nanotubes, when brought into contact with liquid mercury, exhibit not
only even, but also odd multiples of the conductance quantum
$G_0=2e^2/h{\approx}(12.9$~k$\Omega)^{-1}$, whereas the conductance of
individual tubes has been predicted to be exactly $2G_0$
\cite{CNT-conductance}. An even bigger surprise was the observation of
non-integer quantum conductance values, such as $G{\approx}0.5G_0$,
since conductance is believed to be quantized in units of $G_0$
\cite{Landauer}.
 
In this Letter, we demonstrate that the unexpected conductance
behavior can arise from the inter-wall interaction in multi-wall or in
bundled nanotubes. This interaction may not only block some of the
quantum conductance channels, but also redistribute the current
non-uniformly over the individual tubes. We show that under the
experimental conditions described in Ref.~\onlinecite{Heer98}, this
effect may reduce the conductance of the whole system to well below
the expected value of $2G_0$.
 
The electronic band structure of single-wall
\cite{{Mintmire92},{RSaito92},{Oshiyama92}} and multi-wall carbon
nanotubes \cite{{RSaito93},{Lambin94},{PRLxDWT}}, as well as
single-wall nanotube ropes \cite{{Louie98},{PRLxNTR}} is now well
documented. More recently, it has been shown that inter-wall coupling
leads to the formation of pseudo-gaps near the Fermi level in
multi-wall nanotubes \cite{PRLxDWT} and single-wall nanotube ropes
\cite{{Louie98},{PRLxNTR}}. These studies have described infinite
periodic structures, the conductance of which is quantized in units of
$2G_0$. In what follows, we study the effect of inter-wall coupling on
the transport in {\em finite} structures.
 
To determine the conductance of finite multi-wall nanotubes, we
combine a linear combination of atomic orbitals (LCAO) Hamiltonian
with a scattering technique developed recently for magnetic
multilayers \cite{noi1,noi2}. The parametrization of the LCAO matrix
elements, based on {\em ab initio} results for simpler structures
\cite{PRLxCAR}, has been used successfully to describe electronic
structure details and total energy differences in large systems that
were untreatable by {\em ab initio} techniques. This electronic
Hamiltonian had been used previously to explain the electronic
structure and superconducting properties of the doped C$_{60}$ solid
\cite{C60-supercond}, the opening of a pseudo-gap near the Fermi level
in bundled and multi-wall nanotubes \cite{{PRLxNTR},{PRLxDWT}}.
 
The scattering technique, which has recently been employed in studies
of the giant magnetoresistance \cite{noi1,noi2}, determines the
quantum-mechanical scattering matrix $S$ of a phase-coherent
``defective'' region that is connected to ``ideal'' external
reservoirs \cite{noi1}. At zero temperature, the energy-dependent
electrical conductance $G(E)$ is given by the Landauer-B\"uttiker
formula \cite{but85}
\begin{equation}
G(E)=\frac{2e^2}{h}T(E)\;,
\label{la-bu}
\end{equation}
where $T(E)$ is the total transmission coefficient evaluated at the
energy $E$ which, in the case of small bias, is the Fermi energy $E_F$
\cite{S-matrix}.
 
For a homogeneous system, $T(E)$ assumes integer values corresponding
to the total number of open scattering channels at the energy $E$. For
individual $(n,n)$ ``armchair'' tubes, this integer is further
predicted to be even \cite{CNT-conductance}, with a conductance
$G=2G_0$ near the Fermi level. As a reference to previous results
\cite{CNT-conductance}, the density of states and the calculated
conductance of an isolated $(10,10)$ nanotube is shown in
Fig.~\ref{Fig1}.
 
 
 
The corresponding results for the (10,10)@(15,15) double-wall nanotube
\cite{PRLxDWT} and the (5,5)@(10,10)@(15,15) triple-wall nanotube,
where the inter-wall interaction significantly modifies the electronic
states near the Fermi level, are shown in Fig.~\ref{Fig2}. The density
of states of the double- and the triple-wall nanotubes are shown in
Figs.~\ref{Fig2}(a) and (b), respectively. The corresponding results
for the total conductance are given in Figs.~\ref{Fig2}(c) and (d),
respectively. The conductance results suggest that some of the
conduction channels have been blocked close to $E_F$. The inter-wall
interaction, which is responsible for this behavior, also leads to a
redistribution of the total conduction current over the individual
tube walls. The partial conductances of the tube walls are defined
accordingly as projections of the total conductance and shown in
Fig.~\ref{Fig3}. We notice that the partial conductance is strongly
non-uniform within the pseudo-gaps, where the effects of intertube
interactions are stronger.
 
 
 
The experimental set-up of Ref.~\onlinecite{Heer98}, shown
schematically in Fig.~\ref{Fig4}(a), consists of a multi-wall nanotube
that is attached to a gold tip of a Scanning Tunneling Microscope
(STM) and used as an electrode. The STM allows the tube to be immersed
at calibrated depth intervals into liquid mercury, acting as a
counter-electrode. This arrangement allows precise conduction
measurements to be performed on an isolated tube. The experimental
data of Ref.~\onlinecite{Heer98} for the conductance $G$ as a function
of the immersion depth $z$ of the tube, reproduced in Fig.~\ref{Fig5},
suggest that in a finite-length multi-wall nanotube, the conductance
may achieve values as small as $0.5G_0$ or $1G_0$.
 
 
The key problem in explaining these experimental data is that nothing
is known about the internal structure of the multi-wall nanotube or
the nature of the contact between the tube and the Au and Hg
electrodes. We have considered a number of different scenarios and
have concluded that the experimental data can only be explained by
assuming that {\em (i)} current injection from the gold electrode
occurs only into the outermost tube wall, and that {\em (ii)} the
chemical potential equals that of mercury only within the submersed
portion of the tube. In
other words, the number of tube walls in contact with mercury depends
on the immersion depth. The first assumption implies that the
electrical contact between the tube and the gold electrode involves
only the outermost wall, as illustrated in Fig.~\ref{Fig4}(a). The
validity of the second assumption -- in spite of the fact that mercury
only wets the outer tubes -- is justified by the presence of the
inter-wall interaction. The main origin of the anomalous conductance
reduction, to be discussed below, is the backscattering of carriers
at the interface of two regions with different numbers of walls due to
a discontinuous change of the conduction current distribution.
 
The conductance calculation is then reduced to a scattering problem
involving a semi-infinite single-wall nanotube (the one in direct
contact with gold) in contact with a scattering region consisting of
an {\em inhomogeneous} multi-wall tube and the Hg reservoir as the
counter-electrode. Depending on the immersion depth, denoted by
Hg(\#1), Hg(\#2), and Hg(\#3) in Fig.~\ref{Fig4}(a), portions of the
single-wall, the double-wall, and even the triple-wall sections of the
tube are submersed into mercury. Our calculations are performed within
the linear-response regime and assume that the entire submersed
portion of the tube is in ``direct contact'', i.e. equipotential with
the mercury. Increasing the immersion depth from Hg(\#1) to Hg(\#2)
and Hg(\#3), an increasing number of walls achieve ``direct contact''
with mercury, thereby changing the total conductance $G(E)$, as shown
in Figs.~\ref{Fig4}(b)-(d) and Fig.~\ref{Fig5}. We also notice that
the conductance of the inhomogeneous multi-wall structure of
Fig.~\ref{Fig4}(a) cannot exceed that of a single-wall nanotube, which
is the only tube in electrical contact with the gold electrode.
 
The calculation underlying
Fig.~\ref{Fig4}(b) for the submersion depth Hg(\#1) considers a
scattering region consisting of a finite-length (5,5)@(10,10)@(15,15)
nanotube connected to another finite segment of a (10,10)@(15,15)
nanotube. This scattering region is then connected to external
semi-infinite leads consisting of (15,15) nanotubes. The calculation
for the submersion depth Hg(\#2), shown in Fig.~\ref{Fig4}(c),
considers a scattering region formed of a finite-length
(5,5)@(10,10)@(15,15) nanotube segment that is attached to a (15,15)
nanotube lead on one end and to a (10,10)@(15,15) nanotube lead on the
other end. Results in Fig.~\ref{Fig4}(d) for the submersion depth
Hg(\#3) represent the conductance of a (5,5)@(10,10)@(15,15) nanotube
lead in contact with a (15,15) nanotube lead. The calculated
conductances depend on the position of the Fermi level within the
tube. Even though $E_F$ may vary with the immersion depth due to a
changing contact potential, we expect these changes to lie within the
narrow energy window of ${\approx}0.05$~eV, indicated by the shaded
region. Our results of Figs.~\ref{Fig4}(b)-(d), suggesting discrete
conductance increases from $G{\approx}0.5G_0$ for Hg(\#1) to
$G{\approx}1G_0$ for Hg(\#2) and Hg(\#3) are in excellent agreement
with the recent experimental data of Ref.~\cite{Heer98}, presented in
Fig.~\ref{Fig5}.
 
It is essential to point out that from our calculations we expect a
conductance value $G{\approx}0.5G_0$ only when a single tube wall is
in direct contact with mercury. In the case that a single-wall region
is long, we expect this small conductance value to extend over a
large range of immersion depths \cite{Heer98}. In absence of such a
single-wall segment, we expect for the conductance
only values of $1G_0$ and above. We believe that the anomalous
sample-to-sample variation of the observed conductance \cite{Heer98}
is related to the structural properties of the nanotube and not to
defects which are believed to play only a minor role in transport
\cite{Ando}.
 
We also want to point out that a very difference conductance behavior
is expected when more than on tube is in direct contact with the Au
electrode. As a possible scenario, we consider an inhomogeneous
nanotube similar to that in Fig.~\ref{Fig4}(a), where now all of the
three tube walls are in direct contact with the gold electrode. With
two conduction channels per tube wall, the conductance has an upper
bound of $6G_0$. Calculations analogous to those presented in
Fig.~\ref{Fig4} suggest a minimum conductance value $G{\approx}1G_0$
to occur for a finite-length (10,10)@(15,15) tube segment sandwiched
between (15,15) and (5,5)@(10,10)@(15,15) nanotube leads, representing
the smallest submersion depth Hg(\#1), with mercury in direct contact
with only the single-wall portion of the tube. The conductance value
$G{\approx}2G_0$ is obtained for a (10,10)@(15,15) nanotube lead in
contact with a (5,5)@(10,10)@(15,15) nanotube lead, representing
submersion depth Hg(\#2), with mercury in direct contact with a
double-wall tube segment. Finally, depending on the position of $E_F$,
the conductance of a triple-wall nanotube submersed in mercury,
modeled by an infinite (5,5)@(10,10)@(15,15) tube, may achieve
conductance values of $4G_0$ or $6G_0$. Even though the inter-wall
interaction leads to a significant suppression of the conductance, the
predicted increase in the conductance from $1G_0$ to $2G_0$ and $4G_0
- 6G_0$ with increasing submersion depth is much larger than in the
scenario of Fig.~\ref{Fig4}. Also the predicted conductance values are
very different from the experimental data of Ref.~\onlinecite{Heer98},
thus suggesting that only the outermost tube is in electrical contact
with the gold electrode.
 
In conclusion, we have shown that the inter-wall interaction in
multi-wall nanotubes not only blocks certain conduction channels, but
also re-distributes the current non-uniformly across the walls. We
have calculated conductance in several structures and concluded that
the puzzling observation of fractional quantum conductance in
multi-wall nanotubes can be explained by assuming that only the
outermost tube wall is in electrical contact with the gold electrode.
Moreover, we have shown that the sample-to-sample variations in the
conductance are entirely related to the structure of the nanotube. A
very similar behavior is expected for bundled single-wall tubes of
different length submersed into Hg or another metal. These predictions
raise important questions concerning the nature of the nanotube/metal
interface, which deserve further investigation both experimentally and
theoretically.
 
We acknowledge fruitful discussions with Walt de Heer. This work has
been done in collaboration with the group of J.H.~Jefferson at DERA.
SS acknowledges financial support by the DERA and the MSU-CMPT visitor
fund. YKK and DT acknowledge financial support by the Office of Naval
Research under Grant Number N00014-99-1-0252.
 

 
 
 
 
 
\begin{figure}
\caption{
Electronic density of states (DOS) (a) and conductance $G$ (b) of an
isolated single-wall (10,10) carbon nanotube. The DOS is given in
arbitrary units, and $G$ is given in units of the conductance
quantum $G_0=2e^2/h{\approx}(12.9$~k$\Omega)^{-1}$.
}
\label{Fig1}
\end{figure}
 
\begin{figure}
\caption{
Electronic density of states and conductance of a double-wall
(10,10)@(15,15) nanotube [(a) and (c), respectively], and a
triple-wall (5,5)@(10,10)@(15,15) nanotube [(b) and (d),
respectively].
}
\label{Fig2}
\end{figure}
 
\begin{figure}
\caption{
Partial conductance of the constituent tubes of (a) a double-wall
(10,10)@(15,15) nanotube and (b) a triple-wall (5,5)@(10,10)@(15,15)
nanotube. Values for the outermost (15,15) tube are given by the
solid line, for the (10,10) tube by the dashed line, and for the
innermost (5,5) tube by the dotted line.
}
\label{Fig3}
\end{figure}
 
\begin{figure}
\caption{
(a) Schematic geometry of a multi-wall nanotube that is being
immersed into mercury up to different depths labeled Hg(\#1),
Hg(\#2), and Hg(\#3).
Only the outermost tube is considered to be in contact with the
gold STM tip on which it is suspended. The conductance of this
system is given in (b) for the immersion depth Hg(\#1), in (c) for
Hg(\#2), and in (d) for Hg(\#3) as a function of the position of
$E_F$. The Fermi level may shift with changing immersion depth
within a narrow range indicated by the shaded region.
}
\label{Fig4}
\end{figure}
 
\begin{figure}
\caption{
Conductance $G$ of a multi-wall nanotube as a function of immersion
depth $z$ in mercury. Results predicted for the multi-wall nanotube
discussed in Fig.~\protect\ref{Fig4}, given by the dashed line, are
superimposed on the experimental data of Ref.\
\protect\onlinecite{Heer98}.
}
\label{Fig5}
\end{figure}
 
\newpage
 
\begin{figure}
\narrowtext
\epsfxsize=15cm
\centerline{\epsffile{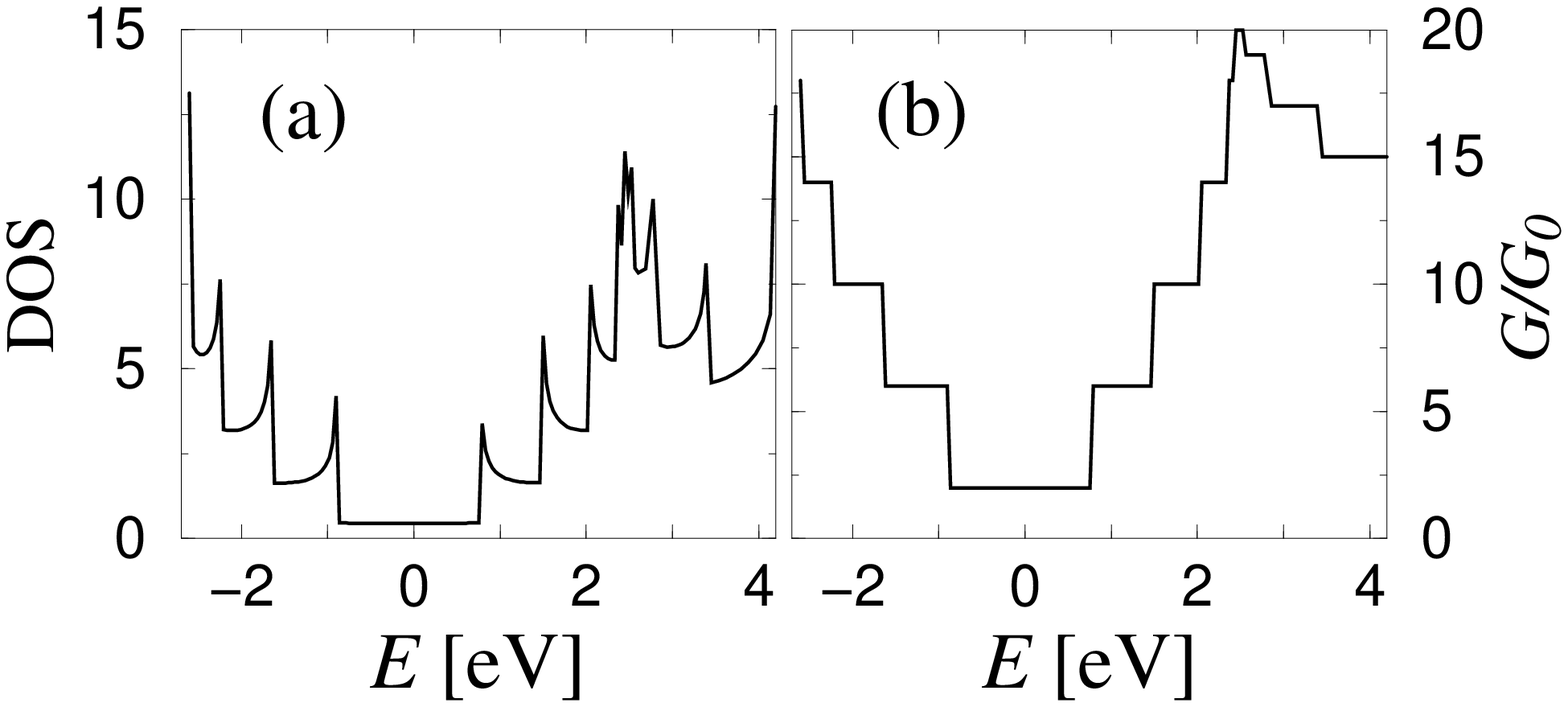}}
\end{figure}
 
\vspace*{1.0cm}
 
\begin{center}
Figure 1 \\
\end{center}
 
\vskip 0.0cm\normalsize
\begin{center}
(Stefano Sanvito, Young-Kyun Kwon,
David Tom\'anek, and Colin J. Lambert,\\
{\em ``Fractional quantum conductance in carbon nanotubes''})
\end{center}
 
\vfill
 
\newpage
 
\begin{figure}
\narrowtext
\epsfxsize=15cm
\centerline{\epsffile{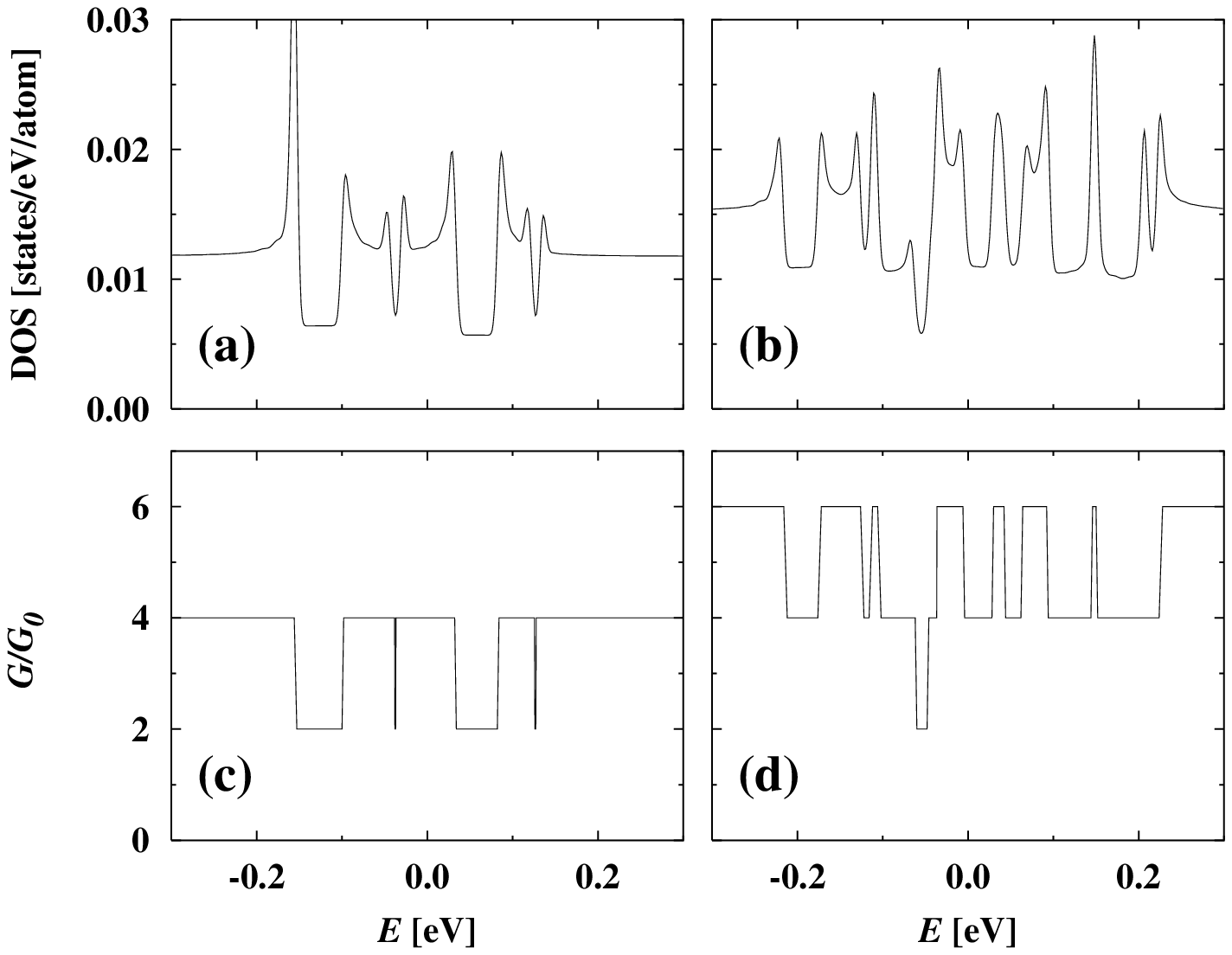}}
\end{figure}
 
\vspace*{1.0cm}
 
\begin{center}
Figure 2 \\
\end{center}
 
\vskip 0.0cm\normalsize
\begin{center}
(Stefano Sanvito, Young-Kyun Kwon,
David Tom\'anek, and Colin J. Lambert,\\
 {\em ``Fractional quantum conductance in carbon nanotubes''})
\end{center}
 
\vfill
 
\newpage
 
\begin{figure}
\narrowtext
\epsfxsize=15cm
\centerline{\epsffile{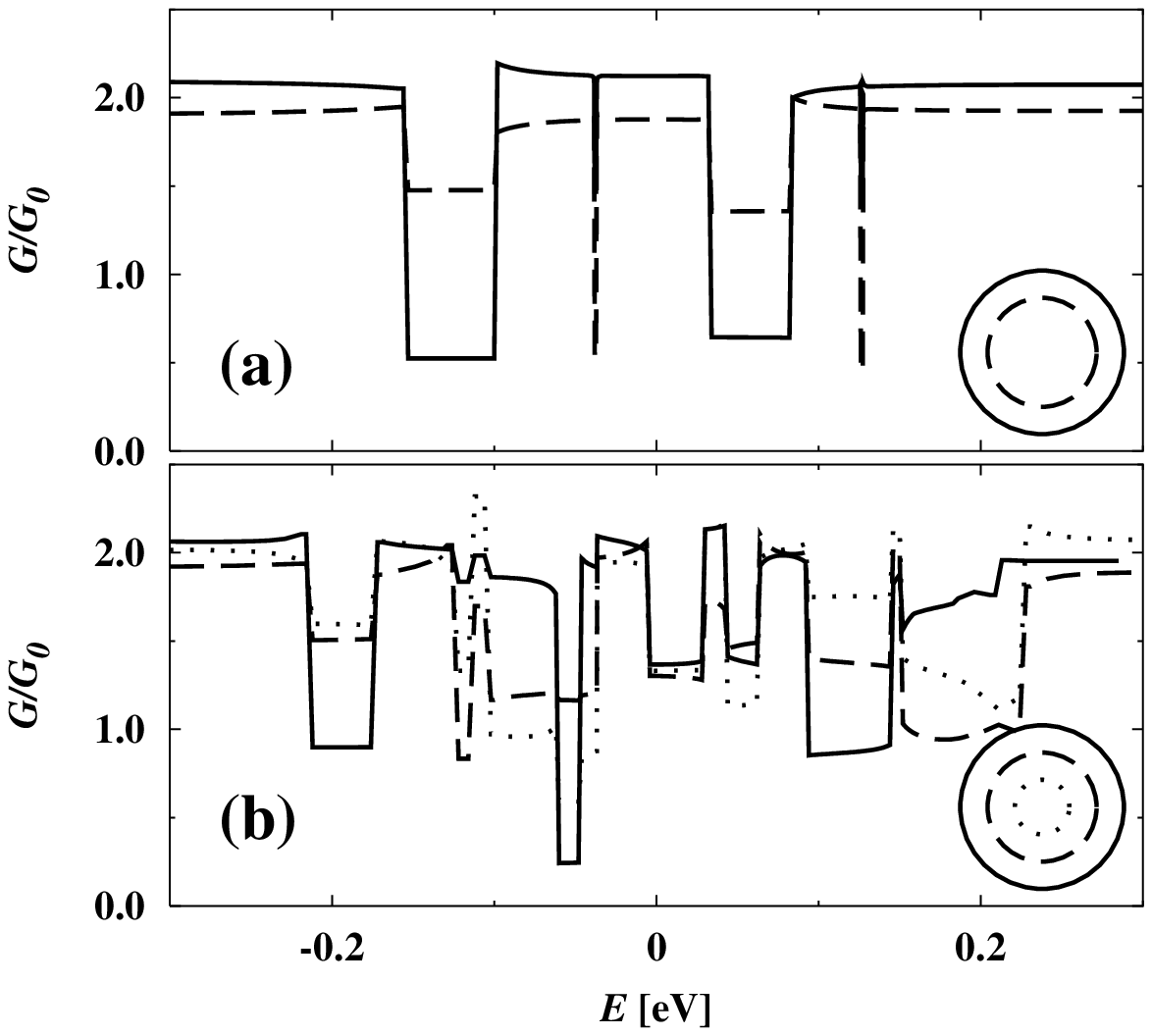}}
\end{figure}
 
\vspace*{1.0cm}
 
\begin{center}
Figure 3 \\
\end{center}
 
\vskip 0.0cm\normalsize
\begin{center}
(Stefano Sanvito, Young-Kyun Kwon,
David Tom\'anek, and Colin J. Lambert,\\
 {\em ``Fractional quantum conductance in carbon nanotubes''})
\end{center}
 
\vfill
 
\newpage
 
\begin{figure}
\narrowtext
\epsfxsize=15cm
\centerline{\epsffile{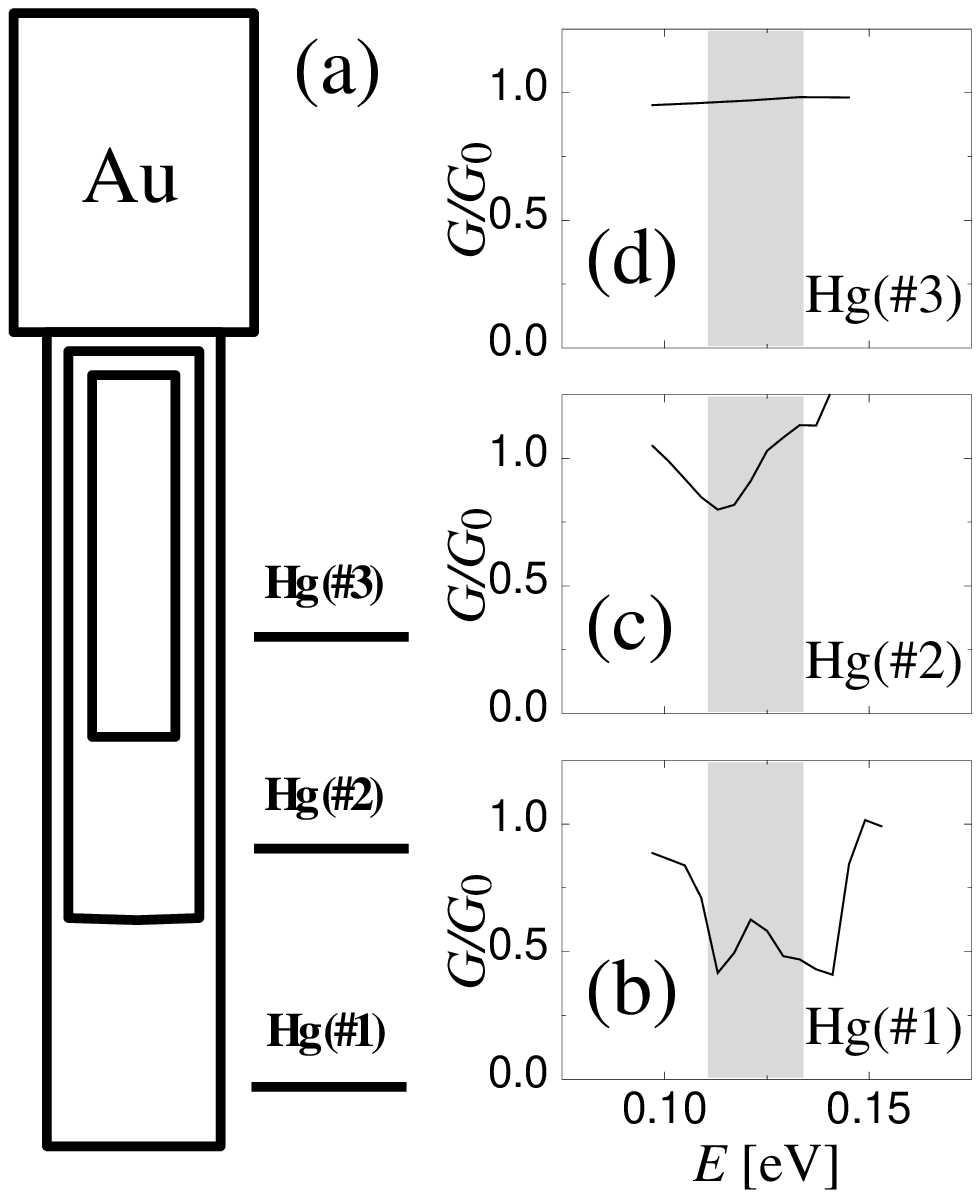}}
\end{figure}
 
\vspace*{1.0cm}
 
\begin{center}
Figure 4 \\
\end{center}
 
\vskip 0.0cm\normalsize
\begin{center}
(Stefano Sanvito, Young-Kyun Kwon,
David Tom\'anek, and Colin J. Lambert,\\
 {\em ``Fractional quantum conductance in carbon nanotubes''})
\end{center}
 
\vfill
 
\newpage
 
\begin{figure}
\narrowtext
\epsfxsize=15cm
\centerline{\epsffile{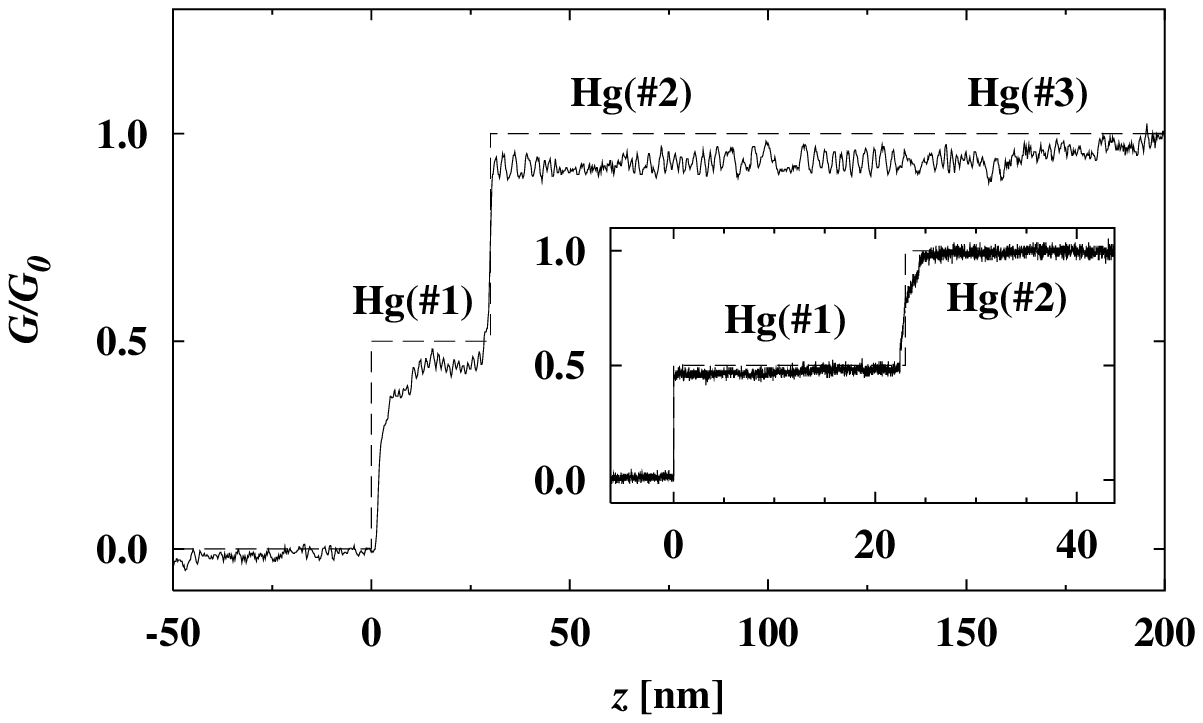}}
\end{figure}
 
\vspace*{1.0cm}
 
\begin{center}
 Figure 5 \\
\end{center}
 
\vskip 0.0cm\normalsize
\begin{center}
(Stefano Sanvito, Young-Kyun Kwon,
David Tom\'anek, and Colin J. Lambert,\\
 {\em ``Fractional quantum conductance in carbon nanotubes''})
\end{center}
 
\vfill
 
 

\begin{references}
\bibitem{S-matrix} When calculating the $S$-matrix of a
         doubly-infinite nanotube, comprising a scattering region
         connected to two semi-infinite single- and multi-wall
         nanotube leads, we note that the unit cell in each segment
         consists of two atomic planes along the nanotube axis. Since
         the hopping matrix between adjacent unit cells ($H_1$ in
         Ref.\ \protect\cite{noi1}) is singular in this case, we
         first project out the non-coupled degrees of freedom before
         calculating the scattering channels.
\end{references}
\end{document}